\documentclass[ reprint, amsmath,amssymb,aps,twocolumn]{revtex4-1}
\usepackage{graphicx}
\usepackage{color}
\usepackage{amsmath}
\usepackage{mathrsfs}
\usepackage{amssymb}
\usepackage{amsthm}
\usepackage{booktabs}
\usepackage[dvipdfm,colorlinks,urlcolor=blue,linkcolor=blue,anchorcolor=blue,citecolor=blue]{hyperref}

\usepackage{lineno}

\begin{document}


\title{Arbitrarily Weak Nonlinearity Can Destroy the Anderson Localization}

\author{Zhen Wang}
\author{Weicheng Fu}
\author{Yong Zhang}
\author{Hong Zhao}
\email{zhaoh@xmu.edu.cn}

\affiliation{Department of Physics and Jiujiang Research Institute, Xiamen University, Xiamen 361005, Fujian, China}

\date{\today }
\begin{abstract}
Whether the Anderson localization can survive from the weak enough nonlinear interaction is still an open question. In this Letter, we study the effect of nonlinear interaction on disordered chain based on the wave turbulence theory. It is found that the equipartition time $T_{eq}$ is inversely proportional to the square of the nonlinearity strength $\lambda$, i.e., $T_{eq}\propto\lambda^{-2}$, in thermodynamic limit.  This result has two fundamentally important consequences. First, the Anderson localized modes can not survive from arbitrarily weak nonlinearity. Secondly, contrary to popular belief, disorder can lead to a more fast thermalization in the weak nonlinear region, which is due to the emergence of three-wave resonance.
\end{abstract}

\maketitle

{\it{Introduction}}.---The Anderson localization \cite{PhysRev.109.1492,abrahams201050} is at the foundation of modern condensed matter physics.  Originally, it is derived for non-interacting disordered systems. However, the interaction is ubiquitous. Therefore whether the localized modes can survive from the weak enough nonlinear interactions is a key question that must be answered. Nevertheless, the localization problem becomes more complex if one goes beyond the picture of noninteracting particles.  Fleishman and Anderson showed that at a low enough temperature electron-electron interaction cannot destroy localization \cite{PhysRevB.21.2366}, while another group found that electron-electron interactions may destroy the constructive interference and lead to a finite electric conductance \cite{lagendijk2009fifty}. In the case of acoustic waves, numerical evidences support either existence \cite{PhysRevLett.102.024101,PhysRevLett.111.064101,PhysRevLett.100.094101,
 PhysRevLett.107.240602,wang2015effects} or absence \cite{PhysRevLett.100.134301,PhysRevLett.100.084103} of a delocalization threshold. This problem is closely correlated to the energy equipartition hypothesis. The hypothesis assumes that arbitrarily weak nonlinear interactions can lead to the energy equipartition among any degree of freedom in the thermodynamic limit. In the case of lattices, it can be expressed as that the energy initially distributed on a small part of modes can eventually share among all of the normal modes. Therefore, if this assumption holds in disordered lattices, the delocalization threshold should not exist.

The numerical verification of the energy equipartition hypothesis is originated from the famous work of Fermi et al. in 1950s \cite{fermi1955studies}.  Since then extensive studies have been carried \cite{campbell2005introduction,berman2005fermi,gallavotti2007fermi,de1999finite,ponno2011two,
izrailev1966stochasticity,casetti1997fermi,Giorgilli2015,
onorato2015route,lvov2018double,pistone2018thermalization,2018arXiv181105697F,2019arXiv190104245F,2018arXiv181208279P,
benettin2011time,benettin2009fermi,fucito1982approach,zhang2016dynamical,ford1992fermi}.  Recently, it is attacked in the framework of wave turbulence (WT) theory \cite{majda1997one,zakharov2012kolmogorov}.  The essential difference of the WT approach to other rigorous approaches \cite{izrailev1966stochasticity,casetti1997fermi,Giorgilli2015} is that it attributes the equipartition to the wave-wave resonance \cite{onorato2015route,lvov2018double,pistone2018thermalization}. The framework contains the following key points. First, there should exist enough sets of resonant waves. Each set contains $p$ waves satisfying the $p$-wave resonance conditions, namely $k_1\pm k_2\pm\cdots\pm k_p \mod N = 0$ for the wave vectors and $\omega_1\pm\omega_2\pm\cdots\pm\omega_p=0$ for the frequencies. Here $k_i$ represents the wave number and $\omega_i$ represents the corresponding frequency of the $i$th normal mode of all $N$ modes. Secondly, resonant sets must be nontrival and  interconnected \cite{onorato2015route}. With these conditions, energy can be transferred among all normal modes and then the equipartition occurs. The scaling of equipartition time $T_{eq}$  is determined by the number $p$ of waves participating in the dominant resonance.  For a typical class of $1$D lattice, characterized by the Hamiltonian
\begin{equation}\label{eq:Hamiltonian}
  H=\sum_{i=1}^{N}\left[\dfrac{p_i^2}{2m_i}+\dfrac{(q_{i+1}-q_{i})^2}{2}+\dfrac{\lambda}{n}(q_{i+1}-q_{i})^n\right], \end{equation}
where $p_i$ and $q_i$ respectively represent the momentum and the displacement from the equilibrium position of the $i$-th atom of mass $m_i$. For convenience of discussion, we rescale Hamiltonian (\ref{eq:Hamiltonian}) with energy density $\epsilon$ by transformation $q_i = q_i'\epsilon^{1/2}$, hence,
the parameter $\lambda$ and  $\epsilon$ has a  rigid scaling relation $\lambda' = \lambda\epsilon^{(n-2)/2}$. Here $\lambda'$ represents the nonlinearity strength, and prime will be omitted for brevity in  the rest of the paper when there is no risk
of confusion. It has been verified \cite{2018arXiv181105697F,2019arXiv190104245F} that the dominant resonances are the $n$-wave ones ($p=n$) for $n\geq 4$ in the thermodynamic limit for $1$D homogeneous lattices, and lead to $T_{eq}\propto\lambda^{-2}$ \cite{2018arXiv181208279P}.
 For the model of $n=3$, three-wave resonances are forbidden due to the dispersion relation, and thus the lowest order resonance in the thermodynamic limit is the four-wave resonance ($p=4$), which leads to $T_{eq}\propto \lambda^{-4}$ \cite{onorato2015route}.

In this Letter, we verify based on the WT theory that  the scaling law $T_{eq}\propto\lambda^{-2}$ can be extended to $n\geq3$ for sufficiently large $1$D disordered chains. We illustrate that this is due to the
removal of the constrain on the wave numbers, as a result of the translation symmetry breaking.
For sufficiently large systems,  the normal frequencies are dense and broadening, and we can expect that the resonance condition of frequency will be satisfied easily without limitation of dispersion relation.
This result has two fundamental consequences. First, the scaling of  $T_{eq}\propto  \lambda^{-2}$  should apply for disordered $1$D lattices with smooth interaction potential. Moreover, three-wave resonances will dominate the long-time dynamics, given that the cubic term is the lowest order nonlinear term of Taylor series of potential. This means that $1$D disordered lattices can be more easily thermalized than homogenous counterparts if nonlinearity is sufficiently weak. Second, there should be no delocalization threshold in $1$D disordered lattices, as a result, the localized modes can not persist in the presence of nonlinear interactions.

{\it{Theoretical analysis}}.---Isotopic disorder enters the model (\ref{eq:Hamiltonian}) via random masses $m_i$ which fluctuates around $\langle m_i\rangle=1$. In the present work, the random masses $m_i$ are chosen independently and identically from a uniform distribution between $1-\delta m$ to $1-\delta m$, the strength of disorder is then characterized by $\delta m$. In general, normal modes of disordered systems can be obtained by diagonalizing the harmonic matrix, which is defined as

\begin{equation}\label{eq:harmonic matrix}
  \mathbf{\Phi}={\Phi_{ij}}=\dfrac{1}{\sqrt{m_im_j}}
\left.\dfrac{\partial ^2H}{\partial q_i\partial q_j}\right|_{\mathbf{q}=\mathbf{0}}.
\end{equation}
There exists a unitary transformation matrix $\mathbf{U}$, whose columns are the normal modes $u^k$, such that
\begin{equation}\label{eq:unitary transformation}
  \mathbf{U}^{\dag}\mathbf{\Phi}\mathbf{U}=\mathbf{\Omega}^2,
\end{equation}
where $\mathbf{\Omega}$ is a diagonal matrix whose elements are the normal mode frequencies, namely $\Omega_{kk}=\omega_k$. Spectral index $k$ is ordered according to the value of frequency, so that $\omega_k\leq \omega_{k+1}$. \par We begin with defining the direct and inverse discrete transform of the $q_j$ variables as \cite{onorato2015route,lvov2018double,pistone2018thermalization,2018arXiv181105697F,2018arXiv181208279P}
\begin{equation}\label{eq:transform}
  \left\{
    \begin{array}{ll}
      Q_k &= \sum_{j}\sqrt{m_j}q_{j}u_{j}^{k}, \\
      q_{j} & =\sum_{k}Q_ku_{j}^{k}/\sqrt{m_j}.
    \end{array}
  \right.
\end{equation}
 We then introduce the complex amplitude of a normal mode $a_{k}(t)$ as
\begin{equation}\label{eq:complex amplitude}
  a_k(t)=\dfrac{1}{\sqrt{2\omega_k}}\left(P_k-i\omega_kQ_k\right),
\end{equation}
where $P_k=\dot{Q}_k$. Substituting Eq.~(\ref{eq:transform}) and (\ref{eq:complex amplitude}) into Eq.~(\ref{eq:Hamiltonian}), then we get the following Hamiltonian,
\begin{align}\label{eq:simple Hamiltonian}
{H}&=\sum_{k}\omega_k{a}_k{a}_k^{*}+\nonumber\\
&\dfrac{\lambda}{n}
\sum_{k_1,\cdots,k_n}{A}_{1,\cdots,n}\prod_{s=1}^{n}\left({a}_{-k_s}^{*}+{a}_{k_s}\right),
\end{align}
where the matrix $A_{1,\cdots,n}$ weights the transfer of energy among modes $k_1,k_2,\cdots,k_n$ and is given precisely by
\begin{equation}\label{eq:A}
  A_{1,\cdots,n}=(-i)^n\tilde{A}_{1,\cdots,n}\prod_{s=1}^{n}\dfrac{\sqrt{2\omega_{k_s}}}{2\omega_{k_s}},
\end{equation}
where
\begin{equation}\label{eq:Atilde}
  \tilde{A}_{1,\cdots,n}=\sum_{j}\prod_{s=1}^{n}
\left(\dfrac{u_{j+1}^{k_s}}{\sqrt{m_{j+1}}}-\dfrac{u_{j}^{k_s}}{\sqrt{m_{j}}}\right).
\end{equation}
Then, the equation of motion for the $k_1$th complex normal mode reduces to
\begin{equation}\label{eq:motion}
  i\dot{a}_{k_1}=\omega_{k_1}a_{k_1}+\lambda\sum
A_{1,\cdots,n}\prod_{s=2}^n\left(a_{-k_s}^{*}+a_{k_s}\right).
\end{equation}
 The equation (\ref{eq:motion}) has a Hamiltonian structure with canonical variables $\{ia_k,a_k^{*}\}$, describing the time evolution of the amplitudes of the normal modes of the system. To evaluate the equipartition time, it is convenient to introduce the wave action spectral density $D_i\delta_{i}^{j}=\langle a_{k_i}a_{k_j}^{*}\rangle$ following the wave resonance approach. We then obtain the $n$-wave kinetic equation in the thermodynamic limit
\begin{equation}\label{eq:kinetic equation}
  \dot{D}_1=\lambda^{2}\int_{-1}^{1}\left|A_{1,\cdots,n}\right|^2\mathfrak{F}(D_{1,n})\delta(\omega_{1,n})dk_2\cdots dk_n,
\end{equation}
where $\mathfrak{F}(D_{1,n})$ is a function of $D_1,D_2,\ldots,D_n$, and $\delta(\omega_{1,n})$ is the shorthand notation of  delta function $\delta(\omega_{k_1}\pm\omega_{k_2}\pm\cdots\pm\omega_{k_n})$.
Following the WT theory, the $T_{eq}$ is inversely proportional to the amplitude of Eq.~(\ref{eq:kinetic equation}) if the integral doesn't vanish, and thus we have
\begin{equation}\label{eq:scaling law}
  T_{eq}\propto \lambda^{-2}.
\end{equation}

To guarantee a non-vanishing integral for $\mathfrak{F}(D_{1,n})$, it should have
\begin{equation}\label{eq:resonant condition1}
  \omega_{k_1}\pm\omega_{k_2}\pm\cdots\pm\omega_{k_n}=0,
\end{equation}
i.e., the $n$-wave resonance condition for the frequencies should be satisfied. This condition is relatively easy to be satisfied for a sufficiently large system. Firstly, the frequencies will become dense when the system tends to infinitely large. Secondly, each frequency will have a certain degree of broadening due to the nonlinearity.
\begin{figure*}[!htbp]
  \centering
  \includegraphics[width=2\columnwidth]{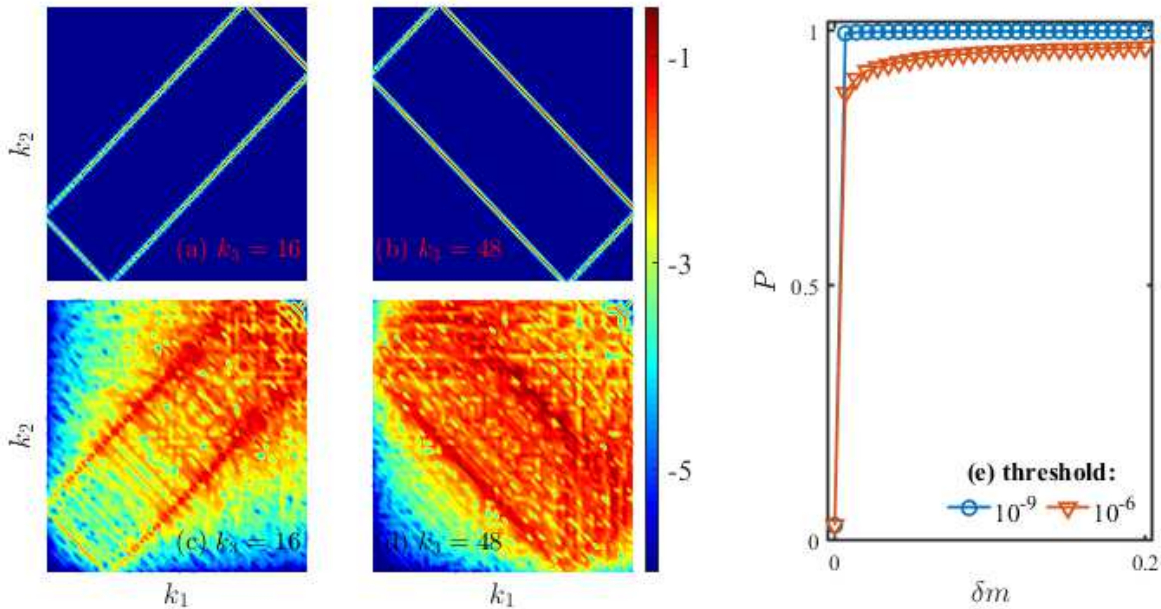}\\
  \caption{The logarithm of $|\tilde{A}_{1,2,3}|$ in $k_1$-$k_2$ planes with fixed $k_3=16$ and $k_3=48$,respectively. (a) and (b) for the homogeneous model, (c) and (d) for the disordered model. (e) The probability of non-vanishing $\tilde{A}_{1,2,3}$ versus the disorder strength.}\label{Figure-1}
\end{figure*}

For the term $A_{1,\cdots,n}$ one can easily show that it reduces to $\delta(k_{1,n})$ multiplying a constant factor if the lattice turns to be homogeneous, which agrees with the result obtained in Ref.~\cite{2018arXiv181105697F}.  In this case, the integral does not vanish when the spectral indexes (the so-called wave numbers in homogeneous lattices) also satisfy the resonance condition, i.e.,
\begin{equation}\label{eq:resonant condition2}
  {k_1}\pm{k_2}\pm\cdots\pm{k_n}\mod N=0.
\end{equation}
Otherwise, by applying the spatial translation invariance and the normal mode solutions to Eq.~(\ref{eq:Atilde}) one can prove that the integral must vanish.  Therefore, to guaranteeing a non-vanishing integral in the homogeneous lattice, one needs to seek for the satisfaction of both resonance conditions (\ref{eq:resonant condition1}) and (\ref{eq:resonant condition2}).  For $n=3$, it is known that these two conditions can not be satisfied simultaneously due to the restriction of the dispersion relation between the spectral index and the frequency.

However, in a disordered system the $\tilde{A}_{1,\cdots,n}$ given by Eq.~(\ref{eq:Atilde}) hardly vanishes due to the absence of the spatial translation invariance.  To confirm this argument, we employ the model of $n=3$ as an example.  In Fig.~\ref{Figure-1}(a) and Fig.~\ref{Figure-1}(b) we plot $|\tilde{A}_{1,2,3}|$ in $k_1$-$k_2$ planes at two fixed $k_3$, i.e., $k_3=16$ and $k_3=48$, respectively, for instance in the case of a homogeneous lattice with $N=64$.   The amplitude of $|\tilde{A}_{1,2,3}|$ is represented by the gray level in the logarithm scale. These plots confirm that $|\tilde{A}_{1,2,3}|$ does not vanish only when the condition (\ref{eq:resonant condition2}) is satisfied . Moreover,  due to the restriction of the dispersion relation, the condition (\ref{eq:resonant condition1}) and (\ref{eq:resonant condition2}) can not be satisfied simultaneously, leading to the vanishing of the integral in Eq.~(\ref{eq:kinetic equation}) and thus forbiddance of the three-wave resonance. However,  by introducing disorder of $\delta m=0.2$ to the lattice,  $\tilde{A}_{1,2,3}$  appears non-zero almost in the entire plane of $k_1$-$k_2$, as shown in Fig.~\ref{Figure-1}(c) and Fig.~\ref{Figure-1}(d). In Fig.~\ref{Figure-1}(e) we further show the probability $P$  of  non-vanishing  $\tilde{A}_{1,2,3}$  as a function of disorder strength $\delta m$ . The probability is calculated by checking the amplitude of $\tilde{A}_{1,2,3}$ for all of the combinations of $k_1$,$k_2$ and $k_3$.  When defining $|\tilde{A}_{1,2,3}| <10^{-9}$ to be the threshold below which it is considered to be vanishing, we find that the probability jumps from $P=0$ to $P=1$ at $\delta m \ne 0$. Even increase the threshold to $|\tilde{A}_{1,2,3}| <10^{-6}$ the jump from $P=0$ to non-zero $P$ is still obvious. Therefore, once the spatial translation invariance is removed, the $\tilde{A}_{1,2,3}$  turns to non-vanishing and Eq.~(\ref{eq:resonant condition1}) remains to be the only one resonance condition.

{\it{Numerical experiments}}.---We apply the numerical simulation to verify that the $n$-wave resonances dominate the thermalization for lattice (\ref{eq:Hamiltonian}) in the thermodynamic limit. Due to the scaling relation between $\epsilon$ and $\lambda$, it is equivalent to studying the scaling
of $\lambda$ by fixing $\epsilon$ or that of $\epsilon$ by fixing $\lambda$. Here, we perform the latter with fixed $\lambda = 1$ for the purpose of verifying Eq.~(\ref{eq:scaling law}), namely $T_{eq}\propto\epsilon^{-(n-2)}$. We adopt the method presented in Ref.~\cite{benettin2011time} to calculate $T_{eq}$. The energy of the $k$th mode is $E_{k}=\left(P_k^2+\omega_k^2Q_k^2\right)/2$. The indicator of thermalization, $\xi(t)=2\tilde{\xi}(t)e^{\eta(t)}/N$, is adopted, where $\eta(t)=-\sum_{k=N/2}^{N}w_{k}(t)\log w_{k}(t)$ is the spectral entropy, in which $w_{k}=E_{k}(t)/\sum_{l=N/2}^{N}E_{l}(t)$, $\tilde{\xi}(t)=2\sum_{N/2}^{N}\bar{E}_{k}(t)/\sum_{1}^{N}E_{k}(t)$, and $\bar{E}_{k}(T)=\frac{1}{(1-\mu)T}\int_{\mu T}^{T}E_k\left(P(t),Q(t)\right)dt$ is the average
energy of the $k$th normal mode. Here, parameter $\mu$  controls the size of the time window for averaging and is fixed at $\mu=2/3$ in our simulation. The equipartition time $T_{eq}$ is measured as that satisfying $\xi(T_{eq})=1/2$.

In simulations, we use the eighth-order Yoshida method \cite{kinoshita1990symplectic} to integrate the equations of motion. The typical integration time step is set to be $\Delta t=0.1$. In order to reduce the fluctuations, as does in Ref.~\cite{benettin2011time}, a further average on $120$ different random choices of initial state is introduced for each realization of disorder. In the following, when there is no risk of confusion we use $\bar{E}_k(t)$ or $\xi(t)$ to denote both variable itself and its average value over $120$ phases.
\begin{figure}[!htbp]
\centering
\includegraphics[width=\linewidth]{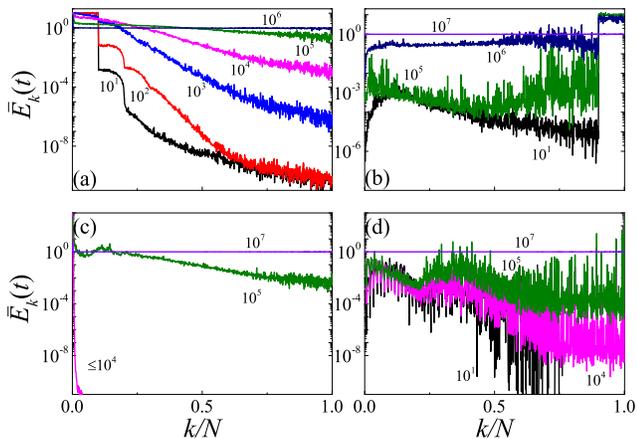}
{\caption{The function $\bar{E}_k(T)$ versus $k/N$ at different times for model of $n=3$ with $N=1023$, $\delta m=0.2$, and $\epsilon=10^{-4}$. Energy is initially distributed among the lowest $10$\% of modes (a), the highest $10$\% of modes (b), the lowest frequency mode (c), and the highest frequency mode (d), respectively. }
\label{Figure-2}}
\end{figure}

Figure \ref{Figure-2} shows the shape of the energy spectrum, more precisely $\bar{E}_k(T)$ vs. $k/N$, at different times. It is for the model of $n=3$ with $N=1023$, $\delta m=0.2$, and $\epsilon=10^{-4}$. Energy is initially distributed among the $10$\% mode of the lowest and highest frequency, respectively in Fig.~\ref{Figure-2}(a) and Fig.~\ref{Figure-2}(b). Note that the normal modes of high frequency are localized while those of low frequency are extended. Despite details in the process are different, we see that in both cases the thermalization eventually occurs. The metastable state, in which the $\bar{E}_k(T)$ keeps its profile nearly unchanged in a very long range of initial time scale and has been found in the homogenous FPUT model \cite{benettin2009fermi} and $\varphi^4$ model \cite{fucito1982approach}, is not found here. This phenomenon is similar to that found in the homogeneous Frenkel-Kontorova model \cite{zhang2016dynamical}. Furthermore,  thermalization can also occur even when only an extended mode or a localized mode is excited.   Figure \ref{Figure-2}(c) and \ref{Figure-2}(d) show respectively the results of energy spectrum initially exciting the mode of lowest (most extended) and the highest (most localized) frequency. Here we emphasize that the energy equipartition begining with localized modes implies the destruction of localization.

\begin{figure}[!htbp]
\centering
\includegraphics[width=\linewidth]{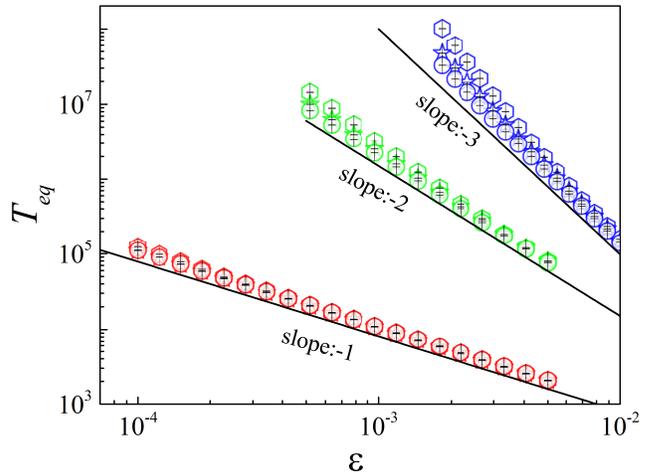}\\
{\caption{The equipartition time $T_{eq}$ as a function of $\epsilon$ in log-log scale for models of $n=3$ (top), $4$ (middle), $5$ (bottom) at system sizes of $N=511$ (hexagon), $1023$ (star), $2047$ (circle). All simulation results are obtain with $\delta m=0.2$. Energy is initially distributed among the lowest 10\% of modes.}
\label{Figure-3}}
\end{figure}

In Fig.~\ref{Figure-3} we show $T_{eq}$  as a function of $\epsilon$ for disordered models of $n=3,4,5$ with $\delta m =0.2$ and different system sizes. It refers to the results when energy is initially distributed among the lowest 10\% of modes. It can be seen that the larger the system size is, the better $T_{eq}\propto\epsilon^{-(n-2)}$ agrees with the data, meanwhile, the lower the energy density is, the larger the size must be to converge to the theoretical prediction. These facts lead us to the conclusion that the prediction keeps correct for arbitrarily low energy density or arbitrarily week nonlinearity in the thermodynamic limit. Meanwhile, for a finite-size system, large deviation from the prediction of slope at low energy density implies the possibility of existence of nonlinearity strength threshold of equipartition.

{\it{Conclusion}}.---We show that $1$D disordered lattices can reach equipartition in the thermodynamical limit. In deed, due to the breaking of the spatial translation invariance, the resonance condition for the wave numbers is removed, at least greatly relaxed, and only the resonance condition for the frequencies of normal modes remains. The resonance will be satisfied more easily between normal modes than that in homogeneous lattices. As a consequence, the universal scaling behavior $T_{eq}\propto\epsilon^{-(n-2)}\lambda^{-2}$, applicable for $n\ge 4$ in homogeneous lattices, can be extended to $n\geq3$ when disorder is involved. Meanwhile, for finite-size systems the scaling law would be violated at sufficiently low energy density.

It is known that the cubic term is the lowest nonlinear term in Taylor series of smooth interaction potential. Therefore, at sufficiently low temperature the cubic term should contribute to the dominant nonlinearity. One can expect that the $T_{eq}\propto\epsilon^{-2}$ works for general $1$D homogeneous lattices while $T_{eq}\propto\epsilon^{-1}$ for general $1$D disordered ones, and reach a conclusion that the disorder can enhance the equipartition instead of inducing the localization of energy at low energy density. Another conclusion of fundamental importance, inferred from this scaling law, is that all of the localized modes will be delocalized eventually with sufficiently small nonlinearity, i.e., the Anderson localization can not survive under the arbitrarily weak nonlinearity in $1$D disordered lattices in thermodynamic limit.

\begin{acknowledgments}
We acknowledge support by NSFC (Grant No. 11335006).
\end{acknowledgments}

\bibliography{reference}

\end{document}